\documentclass[preprint,3p,12pt]{elsarticle}

\usepackage{amsmath}
\usepackage{graphicx}

\usepackage{bm}
\makeatother

\begin{document}

\journal{Journal of Non-crystalline Solids}

\begin{frontmatter}

\title{The effect of inter-cluster interactions on the structure of colloidal clusters}

\author{Alex Malins}

\address{Bristol Centre for Complexity Science, School of Chemistry, University of Bristol, Bristol, BS8
1TS, UK.}

\address{School of Chemistry, University of Bristol, Bristol, BS8
1TS, UK.}

\author{Stephen R. Williams}

\address{Research School of Chemistry, The Australian National University,
Canberra, ACT 0200, Australia.}

\author{Jens Eggers}

\address{School of Mathematics University of Bristol University Walk
Bristol BS8 1TW, UK.}

\author{Hajime Tanaka}

\address{Institute of Industrial Science, University of Tokyo, 
Meguro-ku,Tokyo 153-8505, Japan.}

\author{C. Patrick Royall}

\address{School of Chemistry, University of Bristol, Bristol, BS8
1TS, UK.}

\ead{paddy.royall@bristol.ac.uk}
\date{\today}

\begin{abstract}
Colloidal systems present exciting opportunities to study clusters. Unlike atomic clusters, which are frequently
produced at extremely low density, colloidal clusters may interact with one another.
Here we consider the effect of such interactions on the intra-cluster structure in simulations of colloidal 
cluster fluids. 
A sufficient increase in density leads to a higher population of clusters in the ground state. 
In other words, inter-cluster interactions perturb the intra-cluster behaviour, such that each cluster may no longer be considered as an isolated system.
Conversely, for dilute, weakly interacting cluster fluids little dependence on colloid concentration is observed, and we thus argue that that it is reasonable to treat
each cluster as an isolated system. 
\end{abstract}


\end{frontmatter}

\section{Introduction}

\label{sec:Introduction}

Clusters are a distinct state of matter which exhibit different structural
ordering and phase behaviour, relative to bulk materials~\cite{baletto2005}. Of particular
relevance to, for example, many biological systems such as viruses,
is their tendency to exhibit five-fold symmetry such as icosahedra
and decahedra~\cite{wales}. Recently there has been a surge of interest
in clusters formed in colloidal systems~\cite{segre2001,manoharan2003,stradner2004,campbell2005,sedgwick2004,dibble2006,glotzer2007,wilber2007,zerrouki2008,anthony2008,akcora2009},
leading to the development of `colloidal molecules'
\cite{manoharan2003,glotzer2007,anthony2008,zhang2005diamond,noya2007}.
These may in turn provide novel functionalised materials \cite{manoharan2003,glotzer2007,wilber2007,anthony2008,zhang2005diamond,noya2007}.
In addition to this technological relevance, 
visualisation of colloidal clusters allows direct access 
to their free energy landscape \cite{meng2010}.

Part of the attraction of studying colloidal dispersions is that,
although in principle they are rather complex multicomponent systems,
the spatial and dynamic asymmetry between the colloidal particles
(10 nm-1 $\mu$m) and smaller molecular and ionic species has led to
schemes where the smaller components are formally integrated out 
\cite{likos2001}.
This leads to a one-component picture, where only the \emph{effective}
colloid-colloid interactions need be considered.

Given that the structure of ground state clusters of simple liquid models
is known, along with local energy minima \cite{wales}, it seems natural 
to investigate the prevalence of such structures in colloidal systems,
in particular those with depletion attractions such as colloid-polymer mixtures
where, at fixed real temperature, an effective temperature may be interpreted 
as the inverse of the attraction strength between the colloids.
These colloidal systems 
exhibit similar \cite{mossa2004, arkus2009, wales2010} though not identical \cite{arkus2009,doye1995}
structures to clusters of simple liquid models. 
In general, as the (effective) temperature is reduced, we expect more clusters in the ground state, 
unless kinetic frustration comes into play. 

Unlike atomic clusters, which are often considered in isolation, 
colloidal clusters may themselves form fluids \cite{stradner2004,campbell2005}.
Colloidal cluster fluids are found in systems with competing interactions with 
short-ranged attractions and long-ranged repulsions \cite{stradner2004,campbell2005}. The attractions drive clustering
while the repulsions prevent aggregation and phase separation, leading to a characteristic cluster size \cite{groenewold2001}.
At sufficient concentration, these cluster fluids form gels \cite{campbell2005,sciortino2005bernal,zaccarelli2007,kroy2004},
while for sufficiently strong repulsions, the cluster fluid can undergo dynamical 
arrest \cite{fernandez2009, klix2010} or crystallisation \cite{mladek2006}.
Furthermore, the separation in length- and time-scales allows 
us to treat the system in a hierarchical manner. For example, 
dynamical arrest \emph{within} clusters \cite{malins2009} and of a fluid
comprised of clusters \cite{fernandez2009, klix2010} can in general be decoupled.
This hierarchy means that
one may consider each cluster as an isolated system 
\cite{klix2009}. Alternatively, one may operate at the cluster-cluster level 
\cite{sciortino2004}. Here instead we consider the influence of the \emph{inter-}cluster
interactions on the \emph{intra-}cluster behaviour. In other words,
how is the free energy landscape of a given cluster perturbed by its neighbours?

In a recent experimental study on a cluster fluid \cite{klix2009}, we observed 
that in fact a rather small number of clusters were found in the expected ground state, around 20\%. 
However, after a careful mapping of experimental parameters to conventional
spherically symmetric interactions, Brownian dynamics simulations
showed that isolated clusters reached their ground state, unless geometric
frustration led to ergodicity breaking for deep quenches \cite{malins2009}.

Here we consider an experimentally relevant set of conditions \cite{klix2009}
and use Brownian dynamics simulation to investigate the effects of inter-cluster interactions on the
intra-cluster behaviour. In particular we consider the population of clusters in the ground state.
Our system is characterised by the colloid packing fraction $\phi$, the strength of the attractive interaction $\beta \varepsilon_M$
and the strength of the repulsion $\beta \varepsilon_Y$, where $\beta=\frac{1}{k_{B}T}$, $k_{B}$ is the Boltzmann constant and $T$ is the temperature. We first determine the range of parameters
that correspond to a cluster fluid, before investigating the effect that varying these parameters has on the intra-cluster structure.

\section{Simulation details and model}

Although a great many improvements have since been made \cite{schweizer2002, tuinier2003}, 
the theory of Asakura
and Oosawa (AO) \cite{asakura1954}
is generally accepted to capture the essential behaviour of 
polymer-induced depletion attractions between colloids.
This AO model ascribes an effective
pair interaction between two colloidal hard spheres in a solution
of ideal polymers. However, the hard core of the AO potential
leads to difficulties with Brownian dynamics simulations. The Morse potential
is a variable range spherically symmetric attractive interaction, and
reads

\begin{equation}
\beta u_{M}(r)=\beta \varepsilon_{M}e^{\rho_{0}(\sigma-r)}(e^{\rho_{0}(\sigma-r)}-2),
\label{eq:Morse}
\end{equation}

\noindent where $\rho_{0}$ is a range parameter, $\beta \varepsilon_{M}$ is the potential well depth
and $\sigma$ is the diameter. 

Under the right conditions, this produces similar behaviour to the
Asakura-Oosawa model \cite{taffs2010}. Here we follow our previous
work and set $\rho_{0}=33.06$ and vary the potential well depth $\beta \varepsilon_M$. This
mimics the addition of polymer in the experiments and corresponds to a short-ranged attraction length of $\simeq0.22$ \cite{malins2009}. We used weakly charged
colloids to suppress aggregation. Under some circumstances, one can treat 
the colloid-colloid electrostatic repulsion with a Yukawa form

\begin{equation} 
\beta u_{Y}(r)=\beta \varepsilon_{Y}\frac{\exp(-\kappa(r-\sigma))}{r/\sigma},
\label{eq:Yuk}
\end{equation}

\noindent where $\kappa$ is the inverse
Debye screening length. The contact potential is given by

\begin{equation}
\beta\varepsilon_{Y}=\frac{Z^{2}}{(1+\kappa\sigma/2)^{2}}\frac{l_{B}}{\sigma},
\label{eq:BetaEpsilonYuk}
\end{equation}

\noindent where $Z$ is the colloid charge and $l_{B}$ is the Bjerrum length. 
We use the experimentally relevant value for the inverse Debye length $\kappa\sigma=0.5$ 
and likewise consider two values for the contact potential of the Yukawa repulsion $\beta \varepsilon_Y=1,3$
 \cite{klix2009}. In the experimental
system we seek to model, van der Waals attractions are largely absent. 
Colloids have a hard core, which is often added to equation \ref{eq:Yuk}. Here, however, the Morse potential provides a 
slightly softened core, so we consider the combined potential $\beta u_{M}+\beta u_{Y}$
in the simulations.

For the simulations we use a standard Brownian dynamics simulation scheme~\cite{allen}.
The scheme generates a discrete coordinate trajectory ${\mathbf{r}_{i}}$
as follows\begin{equation}
\mathbf{r}_{i}(t+\delta t)=\mathbf{r}_{i}(t)+\frac{D}{k_{B}T}\sum_{j=1,j\neq i}^{N}\mathbf{F}_{ij}(t)\delta t+\delta\mathbf{r}_{i}^{G},\label{eq:BD}\end{equation}

\noindent where $\delta t$ is the simulation time step, $D$ is the diffusion constant and $N$ is the total number of particles. The colloids respond to
the pairwise interactions $\mathbf{F}_{ij}$ and the solvent-induced thermal
fluctuations $\delta\mathbf{r}_{i}^{G}$ 
are treated as a Gaussian noise with the variance
given by the fluctuation-dissipation theorem. Treatment of the hydrodynamic interaction between colloids is not included in the simulations.

For each state point of $\beta \varepsilon_{Y}=\{1,3\}$, $\phi=\{0.02,0.05,0.10\}$ and 
$\beta \varepsilon_{M}=\{5,6,7,8,9,10,11,12\}$ 
which span the range from monomer fluids to aggregates/gels, 
we perform $8$ statistically independent simulations each of $N=1000$ particles. 
Particles are initialized randomly subject 
to a non-overlap constraint in a cubic box.
Periodic boundary conditions for the box are implemented.
The Morse potential [eq. (\ref{eq:Morse})] 
is truncated and shifted 
for $r>1.5\sigma$, where the Morse potential is typically less than $10^{-6}$.
The electrostatic interactions are treated by adding a Yukawa repulsion
term [eq. (\ref{eq:Yuk})] for different values of $\beta\varepsilon_{Y}$
as specified and this is also truncated and shifted for $r>10.175\sigma$, 
where the potential is of the order $~10^{-3}$.
We study the evolution of the system as the particles condense into small clusters 
under the influence of the Morse attractions, which are stabilized by the long range 
Yukawa charged repulsions.

We define the Brownian time 
as the time taken for a colloid to diffuse its own radius:

\begin{equation}
\tau_{B}=\frac{(\sigma/2)^{2}}{6D}.\label{eq:browntime}\end{equation}

\noindent In the simulations, $\tau_{B}\approx711$ time units, while in the
experiments $\tau_{B}\sim9$ s~\cite{klix2009}. The time-step is $\delta t=0.03$ 
simulation time units and all runs are equilibrated for $1.5\times10^{7}$ 
steps and run for further $1\times10^{7}$ steps. 
The simulation runs therefore correspond to approximately $1000$ Brownian 
times or around $1$ hour, a timescale certainly comparable to experimental work.

We identify two particles as bonded if the separation of the particle
centres is less than $1.25\sigma$ \cite{malins2009}. Having identified
the bond network, we use the Topological Cluster Classification (TCC)
to determine the nature of the cluster~\cite{williams2007}. This
analysis identifies all the shortest path three, four and five membered
rings in the bond network, and identifies clusters from these base units. 
We use the TCC to
find clusters which are global energy minima of the Morse potential 
\cite{doye1995}. We denote the number of particles 
in a cluster by $m$ and follow Doye \emph{et.al.} in terming 
the global energy minima clusters by the number of particles, the
range of the potential and the point group symmetry of the cluster. 
The global energy minimum $m=3,4,5,6,7$ clusters for the Morse potential with $\rho_{0}=33.06$ are
3A $D_{3h}$ triangle, 4A $T_{d}$ tetrahedron, 5A $D_{3h}$ triangular 
bipyramid, 6A $O_{h}$ octahedron, and 7A $D_{5h}$ pentagonal 
bipyramid respectively and the structures are depicted
in Fig. \ref{figM}. For $m\leq7$
there is only one global minimum for all ranges of the Morse potential. For more details, 
see \cite{williams2007}.

\section{Results}

\label{sec:Results}

\subsection{Phase diagram}

\begin{figure}
\begin{centering}
\includegraphics[width=14cm]{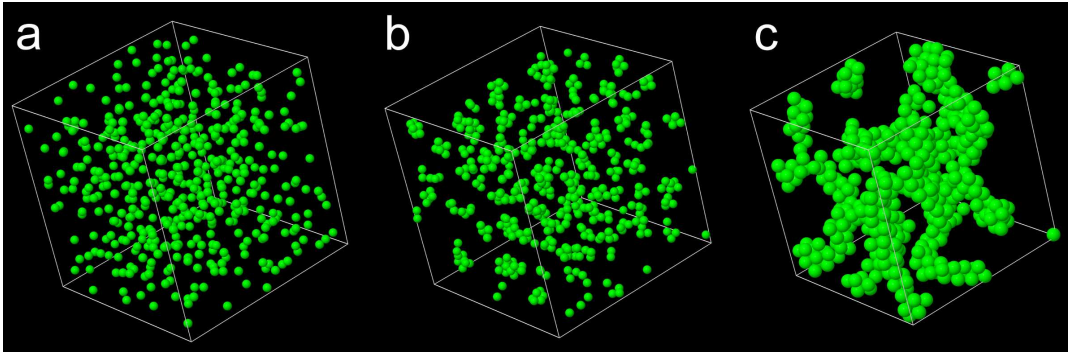}
\par\end{centering}
\caption{Simulation snapshots of the three states seen
(a) $\beta \varepsilon_Y=3, \phi=0.02, \beta \varepsilon_M=5,$ monomer fluid where more than 50\% 
of the particles are identified as monomers/unbonded, 
(b) $\beta \varepsilon_Y=3, \phi=0.02, \beta \varepsilon_M=9,$ cluster fluid in which we are interested,
(c) $\beta \varepsilon_Y=1, \phi=0.10, \beta \varepsilon_M=12,$ aggregate/gel where more than 50\% 
of particles are identified in $m>10$ clusters. }
\label{figShots} 
\end{figure}

\begin{figure}
\begin{centering}
\includegraphics[width=14cm]{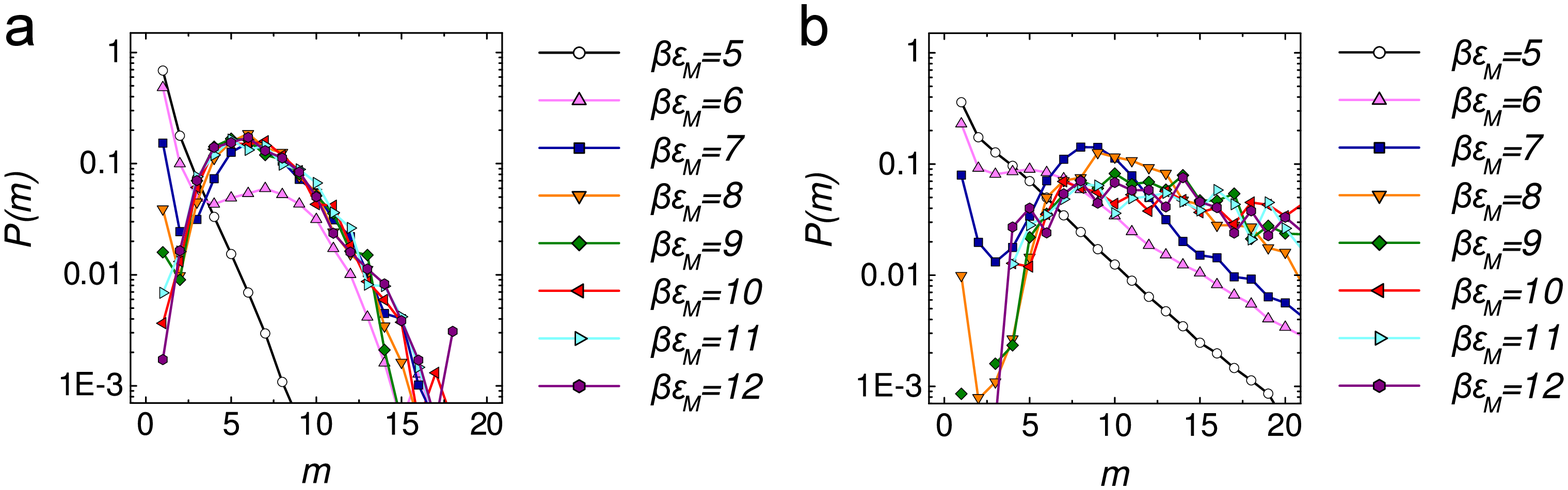}
\par\end{centering}
\caption{Cluster size distributions. $m$ denotes number of particles in a cluster and lines are different values of $\beta \varepsilon_M$.
(a) $\beta \varepsilon_Y=1, \phi=0.02$, (b) $\beta \varepsilon_Y=3, \phi=0.1$.}
\label{figDists} 
\end{figure}


Here we are interested in cluster fluids, however at low density and/or low attraction, the system is dominated
by unbound particles which we term a monomer fluid [Fig. \ref{figShots}(a)] while at high density we find aggregation into much larger,
often elongated \cite{mossa2004,groenewold2001} clusters and ultimately gelation [Fig. \ref{figShots}(c)] \cite{campbell2005, sciortino2005bernal}.
Between the aggregation/gel regime and the monomer fluid lies the cluster fluid seen in Fig. \ref{figShots}(b).

We begin our presentation of the results by determining those state points we consider to be cluster fluids.
To do this, we analyse the cluster size distribution, as shown in Fig. \ref{figDists}. 
In the limit that $\beta \varepsilon_M \rightarrow 0$ we expect a monomer fluid of isolated colloids
as shown in Fig. \ref{figDists}(a) for weakly attracting systems $\beta \varepsilon_Y=1, \phi=0.02, \beta \varepsilon_M \sim 4$.
We find a crossover to a system dominated by clusters at $\beta \varepsilon_M \sim 7$. At higher interaction strengths, the number of monomers drops markedly, as the system moves towards larger clusters.
As Fig. \ref{figDists}(a) shows, for moderate values of the attractive interaction $\beta \varepsilon_M$, 
most clusters have a size $m \leq 10$. We take an arbitrary value of 50\% to distinguish the three states in Fig. \ref{figShots}. In other words,
we consider a system with more than half the particles as monomers to be a monomer fluid, and with more than half the particles in $m>10$ aggregates to be either aggregated or gelled.
Increasing colloid packing fraction $\phi=0.1$ and Yukawa interaction strength $\beta \varepsilon_Y=3$ [Fig. \ref{figDists}(b)] leads to a somewhat different scenario.
Here, we find clusters at all measured attraction strengths. Furthermore, for moderate interaction strengths, the distributions seem rather flat, with
a high incidence of large clusters, or aggregates. In fact no monomers or dimers are present for $\beta \varepsilon_M >8$.
Finding clusters at all interaction strengths is likely related to the increase in packing fraction to $\phi=0.1$ relative to Fig. \ref{figDists}(a). Colloids have less chance to 
avoid one another. Recall the Yukawa repulsion is very long-ranged ($\kappa\sigma=0.5$). The mean interparticle separation $(6\phi/\pi)^{-1/3}$ at $\phi=0.1$ is 
$1.74\sigma$, and the Yukawa repulsion varies only by $1.81 k_BT$ between $1.74\sigma$ and contact, so particles readily approach one another. 
Note that this crossover away from monomers at higher $\beta \varepsilon_M$
is in marked contrast to a number of experiments \cite{stradner2004,campbell2005,sedgwick2004} which found an appreciable number of monomers even at high strengths of the attractive interaction.

\begin{figure}
\begin{centering}
\includegraphics[width=14cm]{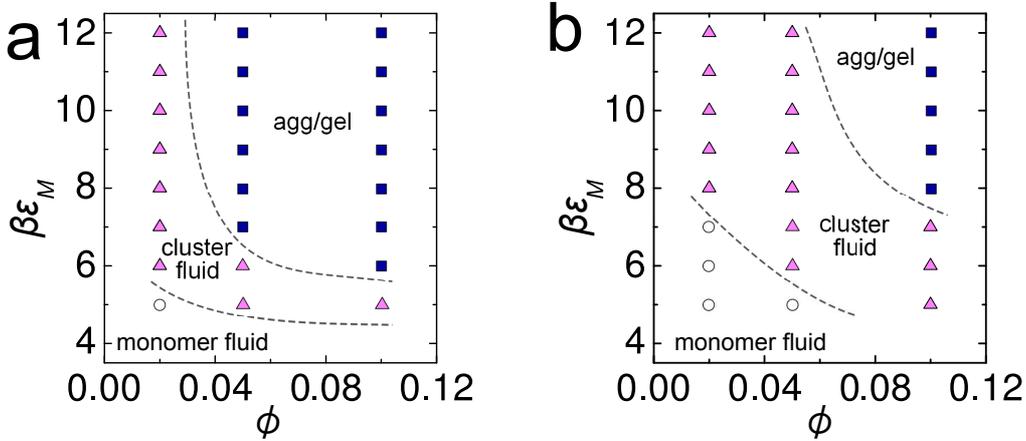} 
\par\end{centering}
\caption{Phase diagrams in the $\phi-\beta \varepsilon_M$ plane. 
States are classified as monomer fluid, cluster fluid or aggregate/gel as defined in the text.
The panels refer to different strengths of the Yukawa repulsion. Dashed lines are a guide to the eye.
(a) $\beta \varepsilon_Y=1.0$,  (b) $\beta \varepsilon_Y=3.0$.}
\label{figPd} 
\end{figure}

Based on the cluster size distributions, we
now determine phase diagrams following our criteria for monomer fluids, cluster fluids and aggregates/gels. These are shown in Fig. \ref{figPd}.
The generic nature is similar to that found in previous simulation \cite{sciortino2005bernal} and experimental \cite{campbell2005} work, of monomer fluids at low density and weak attractions, gels/aggregates at higher density/attraction and a cluster fluid in between. Our definitions lead to a crossover rather than an abrupt transition for both the monomer - cluster fluid and cluster fluid - aggregate/gel. The differences between Figs. 
\ref{figPd}(a) and (b) are reasonably understood in terms of the increase in Yukawa repulsion, which inhibits clustering at the low colloid packing fraction $\phi$ we consider.

\subsection{Inter-cluster interactions}

\begin{figure}
\begin{centering}
\includegraphics[width=14cm]{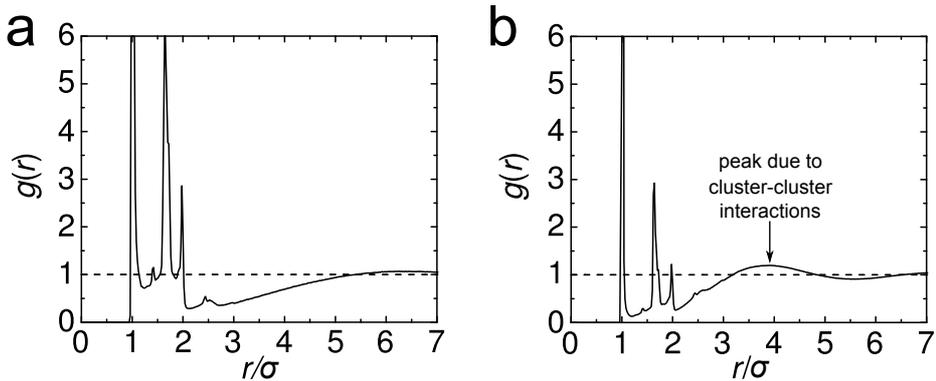} 
\par\end{centering}
\caption{Pair correlation functions in cluster fluids. (a) weakly interacting cluster fluid, $\phi=0.02, \beta 
\varepsilon_Y=1.0, \beta \varepsilon_M=6.0$. (b) strongly interacting cluster fluid, $\phi=0.1, \beta 
\varepsilon_Y=3.0, \beta \varepsilon_M=7.0$ .}
\label{figG} 
\end{figure}

Pair correlation functions are shown in Fig. \ref{figG}. Fig. \ref{figG}(a) is a low-density $\phi=0.02$ $\beta \varepsilon_Y=1.0$ 
$\beta \varepsilon_M=6.0$ system, while
for (b) $\phi=0.1$ $\beta \varepsilon_Y=3.0$ $\beta \varepsilon_M=7.0 $.
In both cases, the intra-cluster structure for $r < 3 \sigma$ shows strong peaks. 
The pair correlation functions also allow us to comment on the strength of the cluster-cluster interactions. Fig. \ref{figG}(b) 
shows a broad peak at $r \sim 4 \sigma$, conversely at longer ranges in Fig. \ref{figG}(a), $g(r)$ is almost flat. The peak in Fig. \ref{figG}(b) 
indicates that these clusters exhibit strong correlations with their neighbours and are thus interacting with one another. Conversely, Fig. \ref{figG}(a) is a weakly interacting system.

By thinking of the cluster fluid as a fluid of (polydisperse) Yukawa-like particles \cite{sciortino2004}, 
we can gain some idea of the interaction between clusters. For the state point in Fig. \ref{figG}(b), the mean cluster size is $<m>\approx8.75$. Typical cluster-cluster separations are given by the cluster-cluster peak i.e. $r_{CC} \sim 4 \sigma$. If we consider clusters of average size at the typical separation, we have a cluster-cluster interaction (i.e. pair interaction) of $\beta U_{CC}=<m>^2 \beta U_Y(r_{CC})\approx 14$. Conversely, in the case of Fig. \ref{figG}(a), there is a very weak peak around $r \approx 6.25 \sigma$, $<m>\approx3.55$, which leads to a typical cluster-cluster interaction strength of $\beta U_Y(r_{CC})\approx 0.15$. We can thus speak of weakly interacting [Fig. \ref{figG}(a)] and strongly interacting [Fig. \ref{figG}(b)] cluster fluids.
We expect that a weakly interacting cluster fluid should show little change relative to isolated clusters, but that in a strongly interacting cluster fluid, the energy landscape of each cluster can be perturbed by its neighbours.

\subsection{Cluster populations}

\begin{figure}
\begin{centering}
\includegraphics[width=14cm]{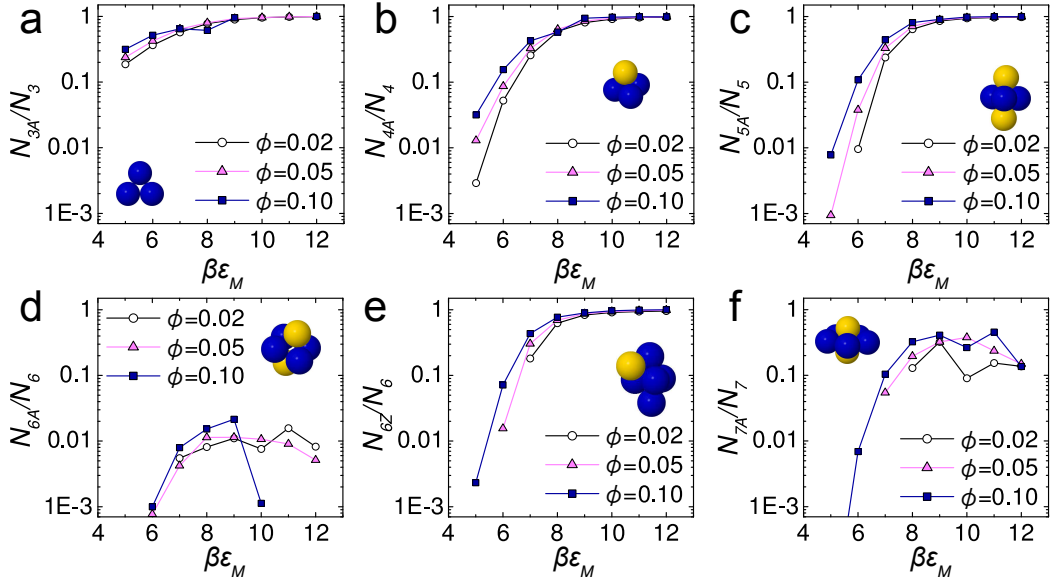} 
\par\end{centering}
\caption{Cluster populations of each cluster size $3 \leq m \leq 7$ as a function
of the well depth of the attractive interaction $\beta\varepsilon_{M}$. Here we fix $\beta \varepsilon_Y=3$.
We define cluster population by the ratio
$N_{c}/N_{m}$ where $N_{m}$ and $N_{c}$ are the total number of clusters of size $m$ and
the total number of clusters of type $c$ (with $m$ colloids) respectively. 
 Symbols refer to colloid packing fraction, $\phi=0.02$, circles,
$\phi=0.05$, triangles, $\phi=0.1$, squares. The different plots corresponding to cluster structures
(a) 3A $D_{3h}$ triangles, (b) 4A $T_{d}$ tetrahedra, (c) 5A $D_{3h}$ triangular bipyramids, (d) 6Z $C_{2v}$, (e) 6A $O_{h}$ octahedra,
 (f) 7A $D_{5h}$ pentagonal bipyramids.}
\label{figM} 
\end{figure}

\begin{figure}
\begin{centering}
\includegraphics[width=14cm]{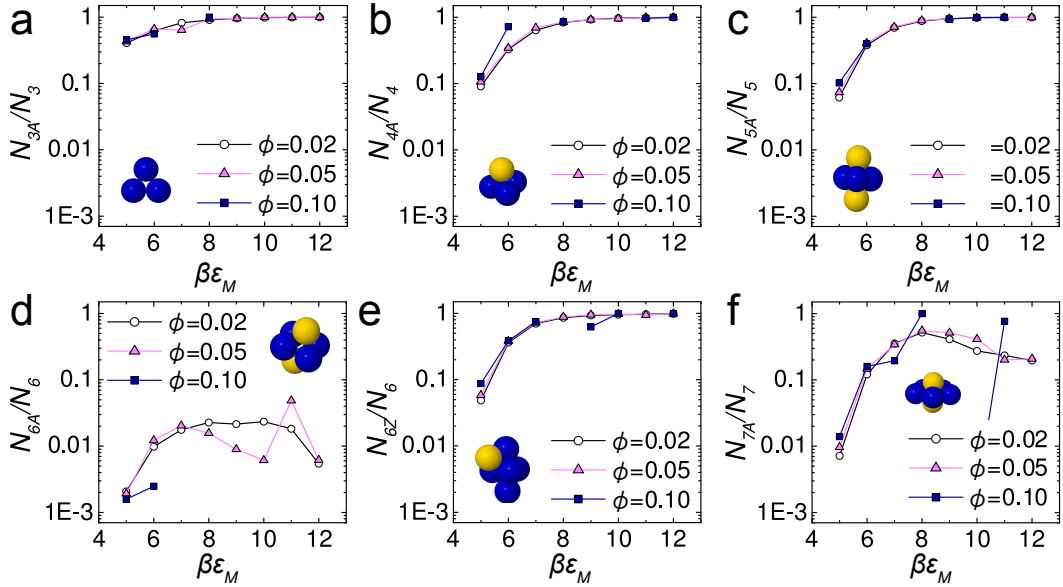} 
\par\end{centering}
\caption{Cluster populations for $\beta \varepsilon_Y=1$. Symbols refer to colloid packing fraction, $\phi=0.02$, circles,
$\phi=0.05$, triangles, $\phi=0.1$, squares. The different plots corresponding to cluster structures
(a) 3A $D_{3h}$ triangles, (b) 4A $T_{d}$ tetrahedra, (c) 5A $D_{3h}$ triangular bipyramids, (d) 6Z $C_{2v}$, (e) 6A $O_{h}$ octahedra,
 (f) 7A $D_{5h}$ pentagonal bipyramids.}
\label{figM2} 
\end{figure}

In the cluster fluid, we treat each cluster individually, and measure its structure. In order to do this we identify the proportion of $m$-membered clusters which are in the ground state. 
We expect that, in the absence of geometric frustration (which is the case for $m<6$),
the cluster population will be dominated by the ground state at sufficient attraction.

Our main results are given in Figs. \ref{figM} and \ref{figM2}. Here we see that, as expected, for $3 \leq m \leq 5$, all clusters at all state points are largely found in the ground state for sufficient values of $\beta \varepsilon_M$. The case of $m=6$  
[Figs. \ref{figM}(d,e) and \ref{figM2}(d,e)]
is somewhat different, as two structures, the 6Z $C_{2v}$ and the 6A $O_{h}$ octahedron compete. Both have the same number of near-neighbours. Without Yukawa repulsions the difference in potential energy for our system is very small (of order 1 part in $1\times10^8$) \cite{malins2009}, which slightly favours the 6A $O_{h}$ octahedron.
However symmetry and vibrational contributions to the free energy favour 
$C_{2v}$ by a factor of around 30 \cite{meng2010} as we indeed see. Here, the inclusion of Yukawa repulsions $\beta \varepsilon_Y>0$ means that 6Z $C_{2v}$ is
somewhat favoured energetically.

In the case of 7-membered clusters [Fig. \ref{figM}(f)], the population of the global minimum 7A $D_{5h}$ pentagonal bipyramid does not tend to unity. 
We showed this is due to ergodicity breaking at relatively deep quenches, such that
geometric frustration can prevent access to the pentagonal bipyramid ground state \cite{malins2009}.
For these relatively short simulation runs ($\tau=400 \tau_B$), once the attractive interaction strength exceeds $\beta \varepsilon_M \sim 5.0$ the bond lifetime exceeds the simulation run time. 


We now turn our attention to the role of increasing density for the $\beta \varepsilon_Y=3.0$ system (Fig. \ref{figM}). We see a general trend of higher density promoting access to the ground state. This is particularly true for those clusters in which there is no geometric frustration limiting access to the ground state, 3A $D_{3h}$ triangles, 4A $T_{d}$ tetrahedra, 5A $D_{3h}$ triangular bipyramids and 6Z $C_{2v}$. 
The effect of colloid packing fraction appears strongest around $\beta \varepsilon_M \sim 6$. For example, we find an order of magnitude increase in the 6Z $C_{2v}$ population at $\beta \varepsilon_M=6.0$ upon increasing the packing fraction from $\phi=0.05$ to $\phi=0.1$ [Fig. \ref{figM}(e)].
For these clusters, 
the effect of raising the colloid concentration may thus be thought of as acting in a similar way to an increase in attraction $\beta \varepsilon_M$. The 6A $O_{h}$ octahedron population is around $1/30$ that of the 6Z $C_{2v}$, which is consistent with simulations \cite{malins2009} and experiments 
\cite{meng2010} on isolated clusters. Compared to isolated systems \cite{malins2009}, we see relatively few 7A $D_{5h}$ pentagonal bipyramids.

In the case of weaker Yukawa repulsions (Fig. \ref{figM2}), our arguments above concerning `strongly' and `weakly' interacting cluster fluids would lead us to imagine that reducing the repulsions would lead to less inter-cluster interactions and less perturbation of the intra-cluster behaviour. This appears to be the case: compared to Fig. \ref{figM}, the relatively weakly interacting clusters in Fig. \ref{figM2} show less response to increasing the colloid packing fraction.
In particular the more dilute $\phi=0.02$ and $0.05$ show little deviation from one another. This suggests that for these parameters, the cluster fluid approaches the dilute limit and behaves as a `cluster gas'. Note that the experimental system considered in \cite{klix2009} was close to $\phi=0.02$, $\beta \varepsilon_Y=1.0$ i.e. `weakly interacting'.

\section{Discussion}

\label{sec:Discussion}

We have considered the effect of cluster-cluster interactions on the 
the intra-cluster structure in Brownian dynamics simulations of a cluster fluid. 
The overall behaviour is broadly similar to isolated clusters. That is to say, 
upon increasing the strength of the attraction
clusters are able to reach their ground states, unless geometrically frustrated from doing so \cite{malins2009}.
For weak cluster-cluster interactions around $0.15 k_BT$ between individual clusters, the intra-cluster behaviour depends little upon density.
We therefore conclude that weakly interacting colloidal cluster fluids at low density (such as $\phi=0.02$, $\beta\varepsilon_Y=1$) may be reasonably treated as independent systems.
Introducing a stronger coupling between the clusters such as 
increasing density leads to a higher population of clusters in the ground state
for a given interaction strength.
In other words, the energy landscape of each cluster is perturbed by its neighbours.

This may be qualitatively understood in terms of the repulsive Yukawa interactions between clusters.
The pair interaction energy between clusters typically
increases as the clusters approach one another upon raising the colloid packing fraction.
These inter-cluster repulsions might be expected to favour `compact' clusters. One example of `compact' is a cluster which minimises its radius of gyration. For spheres, these are 3A $D_{3h}$ triangles, 4A $T_{d}$ tetrahedra, 6A $O_{h}$ octahedra and 7A $D_{5h}$ pentagonal bipyramids \cite{sloane1995}. 
Therefore, increasing cluster-cluster repulsions can lead to similar behaviour as increasing the strength of attraction.
This may be equivalent to noting that clusters of hard spheres which minimize the radius of gyration exhibit the same structures
as for the ground states for particles with attractive interactions 3A $D_{3h}$ triangles, 4A $T_{d}$ tetrahedra, 5A $D_{3h}$ triangular bipyramids, 6A $O_{h}$ octahedra and 7A $D_{5h}$ pentagonal bipyramids
\cite{manoharan2003}. One interesting approach would be to take a system which forms markedly different clusters, such as `patchy particles' 
which can be tailored to form less compact structures in the ground state \cite{wilber2007}. The competition between compact clusters favoured by density and 
the less compact ground state clusters could then be investigated. In this system, there in fact may be some competition between 6Z $C_{2v}$ (favoured in isolation)
and the 6A $O_{h}$ octahedron which is more compact.

It is important to note that, although we simulate on experimentally relevant timescales, the cluster fluids we have studied here may not be truly at equilibrium. Colloidal cluster systems have been shown in simulation to form cluster crystals \cite{mladek2006} or lanes \cite{fornleitner2008}, but this suggests a more uniform cluster size distribution than we find here. Therefore, even though our cluster-cluster interactions can reach $14 k_BT$, the inherent polydispersity in the system prevents crystallisation. 
Systems with competing interactions often exhibit some form of structural ordering \cite{archer2007}. It is reasonable to suppose that much longer simulation runs than we have been able to perform here might result in a shift in the cluster size distribution to one which is more monodisperse. Moreover, state points at higher concentration might be expected to develop into columnar phases or lamellae with a clear periodicity \cite{archer2007,tarzia2006}. 
We have not investigated the cluster dynamics in this system, but it is possible that there is a cluster glass region of the phase diagram.

According to the definitions we have used here, there is no sharp transition between cluster fluids and aggregates/gels. Increasing density/attractions leads to aggregation and gelation. However, locally in the gel we may expect the structure to resemble that of the clusters. This we indeed found in experiments on gelation without long-ranged repulsion \cite{royall2008gel}. With competing interactions, as is the case here, a crossover from a cluster glassy state to a gel has been seen \cite{klix2010}, where gelation is interpreted as a percolation phenomenon. That study found little change in local structure upon gelation. Thus both arrested spinodal type gels without long-ranged repulsions \cite{royall2008gel,lu2008} and those gels resulting from percolation in a system with competing interactions \cite{campbell2005,zaccarelli2007,klix2010} may have a local structure dominated by a topology which follows that of isolated clusters.

\section{Conclusions}

\label{sec:Conclusions}

The effect of cluster-cluster interactions on the intra-cluster structure of a model colloidal system has been studied. 
In the case of weak cluster-cluster interactions at low colloid concentration, the yield of structures is similar to if the clusters were isolated. 
This is the 'cluster gasElimit and interactions between clusters may be neglected. 
By increasing colloid density and/or increasing the strength of the Yukawa repulsion between colloids, cluster-cluster interactions become stronger and may no longer be neglected. 
Increasing either the density or Yukawa repulsion causes a higher yield of ground state structures, showing that energy landscape of a cluster is perturbed by the presence of neighbouring clusters. 
The 'cluster fluidEcannot be approximated as an isolated system due to the presence of cluster-cluster interactions which perturb the intra-cluster structure.

\section*{Acknowledgements}

CPR thanks the Royal Society for funding, AM acknowledges the support of EPSRC grant EP/5011214. This work was carried out using the computational facilities of the Advanced Computing Research Centre, University of Bristol - http://www.bris.ac.uk/acrc/. 

\bibliographystyle{model1-num-names}

\end{document}